# The status and challenges for prostate SBRT treatments in United States proton therapy centers: An NRG Oncology practice survey

*Jiajian Shen, PhD[1], Paige A. Taylor, PhD[2], Carlos E. Vargas, MD[1], Minglei Kang, PhD[3], Jatinder Saini, PhD[4], Jun Zhou, PhD[5], Peilong Wang[1], Wei Liu, PhD[1], Charles B. Simone II, MD[3], Ying Xiao, PhD[6], Liyong Lin, PhD[5]*

[1]*Mayo Clinic Arizona, Phoenix, AZ*

[2]*UT MD Anderson Cancer Center, Houston, TX*

[3]*New York Proton Center, New York, NY*

[4]*University of Washington, Seattle, WA*

[5]*Emory University, Atlanta, GA*

[6]*University of Pennsylvania, Philadelphia, PA*

**Running Title**: NRG Prostate SBRT Survey for Protons

**Number of figures and tables:** 6

**Conflict of Interest Statement**

Liyong Lin, PhD, discloses a grant from Varian Medical Systems, outside the submitted work. Charles B. Simone II, MD, discloses honorarium from Varian Medical Systems, outside the submitted work. Paige A. Taylor, MS, discloses a grant from the National Institutes of Health, during the conduct of the study. The authors have no additional relevant conflicts of interest to disclose.

**Acknowledgment/Funding**

This project was supported by grants U10CA180868 (NRG Oncology Operations), U24CA180803 (IROC) from the National Cancer Institute.

**Data Availability Statement for This Work**

Research data are stored in an institutional repository and will be shared upon request to the corresponding author.

Conceptualization, Data curation, Formal Analysis, Funding acquisition, Investigation, Methodology, Project administration, Resources, Supervision, Validation, Visualization, Writing – original draft, Writing – review and editing

**CRediT (Contributor Roles Taxonomy) Author Statement**

*Jiajian Shen, PhD,* Conceptualization, Formal Analysis, Investigation, Methodology, Validation, Writing – original draft, Writing – review and editing

*Paige A. Taylor, PhD,* Conceptualization, Data curation, Formal Analysis, Investigation, Methodology, Resources, Writing – original draft, Writing – review and editing

*Carlos E. Vargas, MD,* Conceptualization, Formal Analysis, Investigation, Methodology, Supervision, Validation, Writing – original draft, Writing – review and editing

*Minglei Kang, PhD,* Conceptualization, Formal Analysis, Investigation, Methodology, Writing – original draft, Writing – review and editing

*Jatinder Saini, PhD,* Conceptualization, Formal Analysis, Investigation, Methodology, Writing – original draft, Writing – review and editing

*Jun Zhou, PhD,* Conceptualization, Formal Analysis, Investigation, Methodology, Writing – original draft, Writing – review and editing

*Peilong Wang,* Formal Analysis, Investigation, Methodology, Writing – original draft, Writing – review and editing

*Wei Liu, PhD,* Conceptualization, Formal Analysis, Investigation, Methodology, Writing – original draft, Writing – review and editing

*Charles B. Simone II, MD,* Conceptualization, Formal Analysis, Investigation, Methodology, Supervision, Validation, Writing – original draft, Writing – review and editing

*Ying Xiao, PhD*, Conceptualization, Formal Analysis, Investigation, Methodology, Project administration, Resources, Supervision, Writing – original draft, Writing – review and editing

*Liyong Lin, PhD*, Conceptualization, Formal Analysis, Investigation, Methodology, Project administration, Resources, Supervision, Writing – original draft, Writing – review and editing

**Corresponding Author**

Jiajian Shen, PhD

Mayo Clinic Arizona

5881 E Mayo Blvd

Phoenix, AZ 85255

Telephone: (480)-342-0199; Fax: (480)-342-3927

Email: Shen.Jiajian@mayo.edu

**Abstract**

**Purpose:** To report the current practice pattern of the proton stereotactic body radiation therapy (SBRT) for prostate treatments.

**Materials and methods:** A survey was designed to inquire about the practice of proton SBRT treatment for prostate cancer. The survey was distributed to all 30 proton therapy centers in the United States that participate in the National Clinical Trial Network in February, 2023. The survey focused on usage, patient selection criteria, prescriptions, target contours, dose constraints, treatment plan optimization and evaluation methods, patient-specific QA, and IGRT methods.

**Results:** We received responses from 25 centers (83% participation). Only 8 respondent proton centers (32%) reported performing SBRT of the prostate. The remaining 17 centers cited three primary reasons for not offering this treatment: no clinical need, lack of volumetric imaging, and/or lack of clinical evidence. Only 1 center cited the reduction in overall reimbursement as a concern for not offering prostate SBRT. Several common practices among the 8 centers offering SBRT for the prostate were noted, such as using Hydrogel spacers, fiducial markers, and MRI for target delineation. Most proton centers (87.5%) utilized pencil beam scanning (PBS) delivery and completed Imaging and Radiation Oncology Core (IROC) phantom credentialing. Treatment planning typically used parallel opposed lateral beams, and consistent parameters for setup and range uncertainties were used for plan optimization and robustness evaluation. Measurements-based patient-specific QA, beam delivery every other day, fiducial contours for IGRT, and total doses of 35-40 GyRBE were consistent across all centers. However, there was no consensus on the risk levels for patient selection.

**Conclusion:** Prostate SBRT is used in about 1/3 of proton centers in the US. There was a significant consistency in practices among proton centers treating with proton SBRT. It is possible that the adoption of proton SBRT may become more common if proton SBRT is more commonly offered in clinical trials.

Keywords: Stereotactic Body Radiation Therapy, Proton Therapy, Prostate, Clinical Trials, Pencil Beam Scanning

## Introduction

External beam radiation therapy for prostate cancer can span up to nine weeks when employing conventional fractionation. In contrast, only 5 fractions are typically required when delivering stereotactic body radiation therapy (SBRT). The abbreviated treatment course for prostate radiation therapy is supported by radiobiologic experiments indicating that the α/β ratio for the prostate is lower than that of the surrounding tissues[1]. Additionally, numerous clinical trials have shown that moderate hypofractionated[2-4] and SBRT[5-7] treatments for prostate cancer have similar treatment outcomes and toxicity rates as conventional fractionation. Consequently, the utilization of SBRT for prostate cancer has grown over time due to its added convenience for patients and its cost reduction compared to conventional fractionation[8].

In theory, the benefits of SBRT for the prostate should also extend to proton radiation therapy. Such shorter proton courses for prostate cancer can help reduce the current disparities in access to proton therapy[9]. Furthermore, since proton therapy is a valuable and currently still more limited resource, SBRT for prostate could free up machine time, enabling clinics to treat more patients across various disease sites. However, in practice, SBRT for the prostate is not widely adopted in proton therapy. Experiences and outcomes of using protons for prostate SBRT have only been reported in a few papers to date[10-13]. Santos et al[13] presented the 5-year results of 38Gy proton SBRT over 5tx compared to 79.2Gy in 44tx. Freedom from failure, disease free survival and overall survival were similar for both groups. Toxicity was very low in both groups.

In 2021, the NRG Oncology Medical Physics Subcommittee established a working group to investigate the status of using SBRT treatment in proton radiation therapy for the disease sites of lung, liver, spine, and prostate. The working group consisted of radiation oncologists and therapeutic medical physicists with expertise in proton therapy. A survey following European and NRG Oncology precedents[14-17] was designed by the workgroup for each disease site and was distributed to all 30 proton centers in the United States who are members of the National Clinical Trial Network (NCTN).

This paper reports the survey results and summarize the treatment methods for proton centers that have adopted SBRT for prostate cancer. This report can serve as guidelines and references for other proton centers interested in implementing SBRT. Another goal of this analysis was to gather information on the concerns that have impeded some centers from implementing SBRT for prostate so that we can provide potential future directions for the proton industry and clinical teams to address.

## Materials and Methods

The survey was designed by a team of radiation oncologists and therapeutic medical physicists with expertise in proton therapy and clinical trials. The survey included 54 questions across 5 categories. Questions 1-4 focused on gathering basic information, including the institution's name, proton system vendor, beam delivery mode, and the treatment planning system. Questions 5-11 aimed to determine the types of patients selected. Questions 12-16 sought information on CT Simulation, the immobilization method, and fiducial markers. Questions 17-45 delved into prescription, treatment planning processes, and patient-specific QA. Questions 46-54 covered IGRT during treatment and the verification CT. The complete set of survey questions is included in Appendix 1. Responses were either selected from a dropdown menu with pre-populated options (e.g., always, sometimes, never, etc.), or filled in a comment section, or both, so that respondents could provide further explanation of their selection.

The survey was distributed by the Imaging and Radiation Oncology Core (IROC), which monitors proton therapy centers participating in NCTN protocols. It was distributed on February 1$^{st}$, 2023, to all 30 proton therapy centers that participate in NCTN clinical trials in the United States. Several follow up requests were sent to institutions that did not initially respond. Answers were summarized per center in an open manner.

For centers that reported not performing proton SBRT for prostate patients, a follow-up survey with three additional questions was circulated on 05/03/2023 to gather insights into impediments that prevent them from offering this therapy. These follow-up questions are provided in Appendix 2.

## Results

### Availability of Prostate SBRT in Proton Centers

In total, 25 centers responded, representing an 83% participation rate. 8 centers (32%) reported offering SBRT for the prostate using proton therapy, whereas the remaining 17 centers (68%) reported not offering SBRT for the prostate. The three reasons cited by these 17 centers for not adopting SBRT for prostate in proton therapy, which included no clinical need (n=4), hardware limitation (n=6), and lack of clinical evidence (n=7).

### General Information of the Proton System and Patient Selection

Figure 1 displays the general information regarding proton systems and patient selection at the 8 proton centers that employed proton SBRT for prostate treatment. There is no particular predilection for proton vendors or treatment planning system selections. Pencil beam scanning (PBS) delivery mode was used by 7 centers, and the IROC prostate phantom credentialing was completed by 7 centers. Regarding risk levels, 1 center allowed for only low-risk cases, 4 treated low- and intermediate-risk patients, and 3 offer proton SBRT for low-, intermediate-, and high-risk cases. Patients with hip prostheses were not treated in 3 center, 4 centers allowed SBRT for patients with unilateral hip prostheses, and only 1 treated patients with bilateral hip prostheses using SBRT. Cone beam CT for IGRT was used in 2 centers; kV/kV was used in 4 centers; and the other 2 centers used both kV/kV and cone beam CT. The treatment intent across all centers was curative, with only 2 centers employing palliative SBRT treatment for bone metastases. The most common reasons for selecting SBRT for prostate treatment were patient convenience and patient requests. Most centers (5 out of 8) did not utilize SBRT for prostate metastases, and half (4 out of 8) employed SBRT for reirradiation.

**CT simulation and Immobilization**

Table 1 presents information on CT simulation immobilization devices and fiducial makers. Vac-Lok and knee cushions were each used for immobilization in 4 centers. Only 1 center employed an endorectal balloon filled with water. Hydrogel spacers and fiducial markers were available in all centers. Fiducial markers made of carbon and gold were evenly used by 4 centers each. The CT slice thickness varied from 1.0 to 3.0 mm, with all but 1 center preferentially using a slice thinking of 1.0 to 2.0 mm.

**Prescriptions, Contours, Beam Angles and Dose Constraints**

Table 2 provides information on prescriptions, target contours, and beam arrangements. All 8 centers employed 5 fractions delivered every other day. The total dose ranged from 35 to 40 GyRBE. MRI was commonly used to aid in target delineation. PTVs were utilized in 5 centers for plan optimization and evaluation. The margins from CTV to PTV included range uncertainty in the beam direction in 3 centers. In 2 centers, PTV was cropped when OAR dose constraints could not be met. The typical beam arrangements comprised of 2 parallel opposed lateral beams employed by 6 centers. In comparison, 2 centers use 4 beams by supplementing parallel opposed beams with 2 anterior oblique beams with 30% weighting.

Supplementary Table 1 provides the dose constraints reported by 8 centers. Diversities were observed for the target and OAR dose constraints. For instance, the minimum percent volume of CTV should be covered by the prescription doses was 95% for 2 centers, and 98% for another 2 centers.

**Treatment plan optimization and robustness evaluation**

Table 3 presents the methods and parameters for treatment plan optimization and robustness evaluations. Since the center with uniform scanning does not employ inverse plan optimization, 7 centers offering PBS were included in these data. Single field optimization (SFO) is employed in all 7 PBS-treating centers, with 2 centers occasionally using multi-fields optimization (MFO).

Regarding plan optimization, two methods were utilized. The first, used by 2 centers, involved generating an expanded structure around the CTV that encompassed both setup and range uncertainties and then optimizing the plan to cover this expanded structure. The second method, which optimizes the dose to the CTV while directly incorporating setup and range uncertainties during the optimization, was used by 5 centers. Setup margins of 3 mm were employed for the posterior direction, and 3-5 mm margins were used for the other directions. Range uncertainty of 3-3.5% was used. For robustness evaluation, the worst-case scenario for CTV coverage was used.

**Dose Calculation Parameters, 2nd MU check and Patient-Specific QA**

Table 4 presents information on dose calculation, second MU check, and patient-specific QA. Analytical methods for dose calculation were used in 4 centers, whereas the remaining 4 employed Monte Carlo (MC) methods. Statistical uncertainties for doses less than 0.5% to 2% were applied for the MC algorithms. For the analytical algorithms, grid sizes ranged from 2 to 2.5 mm. Monte Carlo was used for 2nd MU check in 3 centers, and analytical algorithms was used in 2 centers. Patient-specific QA was predominately based on measurements, except for 1 center that employed logfile QA for all patients, with measurements applied only to randomly selected patients. The number of measurement planes for patient QA varied from 1 to 3. The centers that performed log file analysis conducted it before the first treatment session, and none of the clinics conducted these analyses for daily treatment monitoring.

**IGRT used in patient alignment and verification CT for dose evaluation**

Table 5 presents the IGRT methods for patient alignment and the verification CTs for dose evaluation. None of the centers employ specific beam rescanning for beam delivery. Fiducial marker contours are utilized for IGRT, with IGRT tolerance varying from 1 to 3 mm. IGRT with CBCT was used by 4 centers, whereas the other 4 centers rely on kV/kV. kV imaging is performed for each field at 5 centers, and only 1 center taking post-treatment kV images. Verification CT before treatment was consistently performed in 2 centers, whereas 3 centers do so when significant deviations are observed in patient setup. Additionally, 5 centers occasionally evaluate the dose using verification CT when necessary.

### Concerns about not implementing proton SBRT for prostate

Among centers who do not currently deliver prostate proton SBRT, 12 proton centers (71%) responded to the follow-up short survey. Among these, 7 centers also do not offer SBRT for the prostate using photon therapy or Brachytherapy. For the remaining 5 centers that treat the prostate with SBRT using photons or brachytherapy, SBRT was not considered for proton therapy due to reasons such as preserving proton capacities for other disease sites, a lack of evidence supporting proton therapy for this indication, and the absence of CBCT in proton facilities.

It was responded by 10 centers that a lower reimbursement rate for proton SBRT was not a concern for not using it for prostate cancer. In contrast, only 1 center mentioned that it was a primary concern, whereas another stated that it was a concern but not the primary one.

While all proton centers acknowledged the convenience of SBRT for prostate patients, its implementation for protons is limited. Regarding why relying solely on kV images and fiducials for prostate SBRT is inadequate, responses indicated concerns such as target deformation, daily variations in the rectum and bladder, and changes in patient habitus that cannot be adequately assessed without volumetric imaging. Concerning the lack of clinical evidence, all responses indicated that SBRT for prostate would be considered if more clinic trials supported its use.

## Discussion

Proton therapy systems involve many vendors offering different treatment modalities and online imaging methods. Furthermore, protons are more sensitive to setup and range uncertainties than photon therapy. Many details concerning proton therapy, such as plan optimization and evaluation, must be considered. Therefore, the NRG Oncology Medical Physics Subcommittee along with the NRG Oncology Particle Therapy Work Group formed this working group to survey the operational proton centers in the United States regarding their practice for prostate SBRT. This survey can help determine whether these practices align with standard guidelines, which can aid in designing future cooperative group and other clinical trials, while also helping proton centers interested in implementing this treatment procedure.

Proton therapy is a relatively scarce resource associated with high delivery costs, and in some cases, it increases patients' travel expenses due to the reduced availability. One significant finding from this survey is that SBRT for prostate cancer is not widely utilized in proton therapy despite its convenience and cost-effectiveness for patients. The responses indicated that the lack of clinical evidence is the primary concern for not adopting it. Figure 1 shows no consensus exists on which patients risk levels should most optimally be treated with SBRT. Clinical trials for protons are essential for comparing them to photons, for standardization, and to have safe and consistent patient treatments. Proton centers should consider conducting more clinical trials with the involvement of more institutions and longer-term follow-up to quantify its efficacy and safety more fully. As case rates have not been adopted in the U.S., more protracted treatments, especially with protons, may have higher reimbursement. Most centers did not mention lower reimbursement as an issue for adopting proton SBRT, but we are cognizant that conversations about reimbursement and fractionation happen on an ongoing basis in many proton centers in the U.S. We are hopeful that reimbursement is not a deterrent to adopting more hypofractionated schedules.

In addition to the limited clinical evidence, the absence of volumetric imaging is a big concern. Compared to modern linacs, where CBCT is a standard technique, CBCT is a more recent addition to proton facilities[18]. Many old proton therapy systems lack volumetric imaging, but some centers are either upgrading or planning to upgrade to CBCT. That would help these clinics implement SBRT. Proton vendors should make the volumetric imaging system as the standard option. With technological advances in online imaging, proton therapy can achieve its full clinical potential and may prove to be a valuable treatment modality when delivering prostate SBRT[19].

Among 8 proton centers that offer SBRT for the prostate, there are many areas of consensus, as demonstrated in the results section. Most proton centers (87.5%) employ PBS delivery mode and completed IROC phantom credentialing. Hydrogel spacers, fiducial markers, and MRI usage are part of clinical practice across all centers. Treatment planning typically involves parallel opposed beams, and consistent parameters for setup and range uncertainties are used for plan optimization and robustness evaluation. Measurements-based patient-specific QA, every-other-day beam delivery, and fiducial contouring for IGRT are consistent practices across all centers. Doses among centers were very uniform with all centers offering doses between 35Gy-40GyRBE in 5 treatments every other day.

Given the relatively small number of proton centers available and the limited number of proton-specific randomized trials conducted, it has been customary to adopt doses and fractionations used in conventional radiation therapy. In this manner, doses and fractionations for most disease sites currently treated with proton therapy have been derived from studies conducted with conventional radiation. Many papers on proton therapy have shown that similar fractionation can provide comparable control and toxicity rates when incorporating a small RBE correction. For prostate SBRT, three randomized trials have consistently demonstrated safety and efficacy. In the Pace B study[6], 36.25 Gy administered over 5 treatments was compared to 78 Gy in 39 treatments or 62 Gy in 20 treatments. Daily image guidance with prostate image-guided radiation therapy was mandatory, and the use of fiducial markers for guidance was permitted. Different margins were employed for SBRT compared to the standard arm, as well as higher hotspots within the target, delivering up to 45 Gy. In the HYPO-RT-PC study[7], 42 Gy over 7 treatments was compared to 78 Gy in 39 treatments. Fiducial marker guidance was used for most patients, and treatment planning and margins were similar between the two groups. In the PCG GU002 study[13], 38 GyRBE over 5 treatments was compared to 79.2 GyRBE over 44 treatments delivered with proton therapy (ASTRO 2023). Daily fiducial guidance, margins, and planning evaluation were consistent for both groups. Among the three studies, it is evident that long-term toxicity rates, quality of life (QOL), and efficacy are similar between the two treatment approaches. These studies provide clinical evidence for the safety and efficacy of an SBRT approach, whether using x-rays or proton therapy, based on fiducial markers.

Our study has some limitations, including the fact that questions and answers were not anonymized, which may have influenced some of the responses. For the technical parameters, we relied on information provided by the center, and treatment data was not directly reviewed, which means that variations in treatment may be larger than reported. Only centers within the NRG group were surveyed, which comprises most radiation centers in the US, but not all. However, our study's strengths include a very high response rate, detailed information about treatment parameters, and subjective assessments of SBRT. This study remains the most in-depth analysis of current proton SBRT practices in the US.

In summary, the survey results revealed that SBRT for prostate is not widely used in proton centers currently. Among the 8 proton centers that offer SBRT for the prostate, several standard practices were reported. It is possible that higher patient demand may drive its future usage.

Figure Legends:

Figure 1. Proton system vendors, beam delivery techniques, treatment planning systems, IROC credentialing, risk level and hip prosthesis for patient selections, and IGRT methods from the 8 proton centers treating prostate cancer with SBRT.

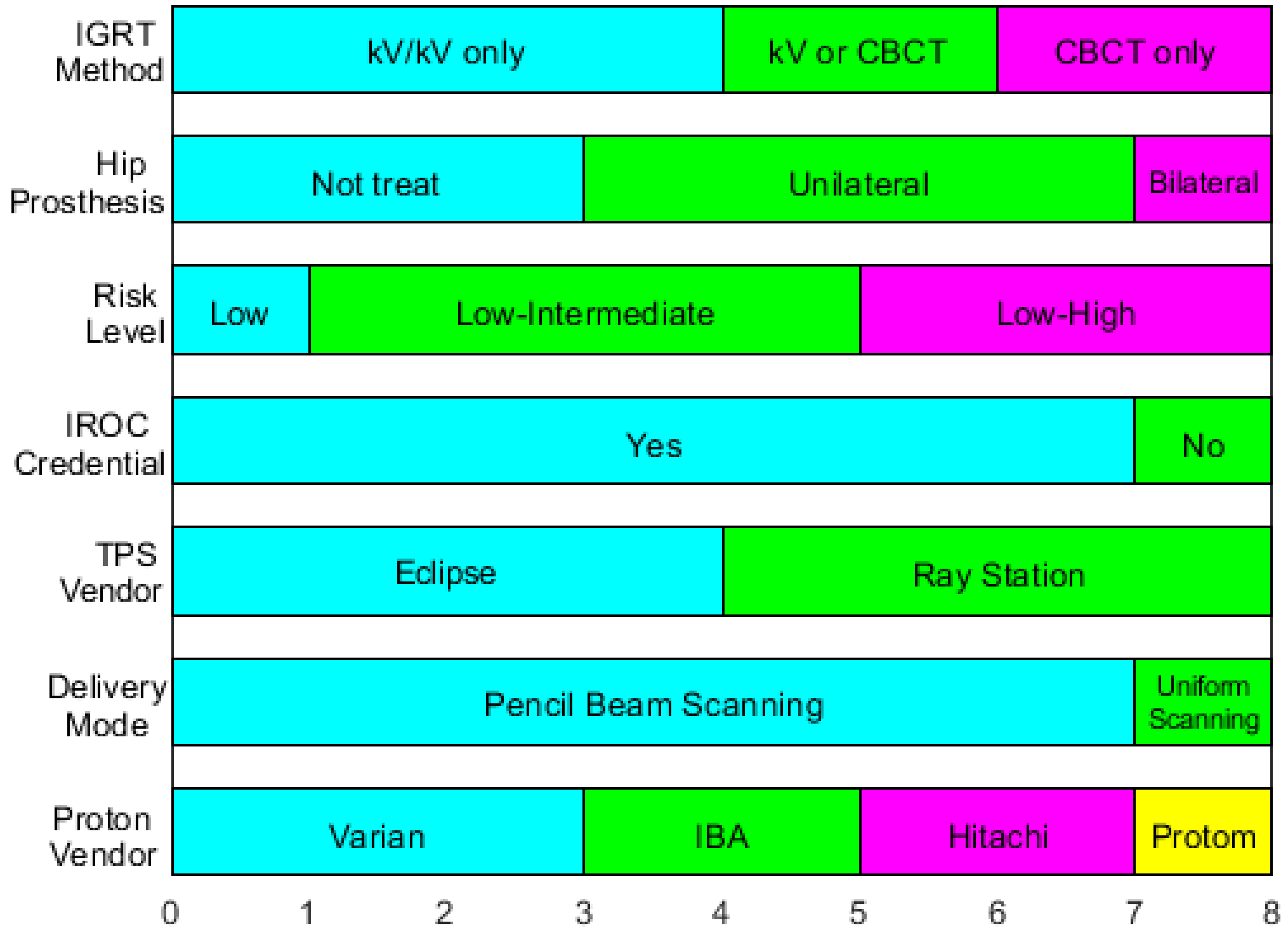

Table 1. CT Sim and immobilization methods for prostate SBRT in proton.

| | Immobilization Devices | Endorectal Balloon | Material in Balloon | Hydrogel Spacer | Fiducial Markers | Marker Material | CT Slice (mm) |
|---|---|---|---|---|---|---|---|
| Center 1 | Vac-Lok | No | N/A | Yes | Yes | Carbon | 1 |
| Center 2 | Knee Cushion | No | N/A | Yes | Yes | Carbon | 2 |
| Center 3 | Knee Cushion | No | N/A | Yes | Yes | Carbon | 1.5 |
| Center 4 | Vac-Lok | No | N/A | Yes | Yes | Gold | 3 |
| Center 5 | Knee Cushion | Yes | Water | Yes | Yes | Carbon | 2 |
| Center 6 | Vac-Lok | No | N/A | Yes | Yes | Gold | 2-3 |
| Center 7 | Knee Cushion | No | N/A | Yes | Yes | Gold | 1.25 |
| Center 8 | Vac-Lok | No | N/A | Yes | Yes | Gold | 1.5 |

Table 2. Prescriptions, target contours, beam angles for the prostate SBRT treatment by protons.

| | Center 1 | Center 2 | Center 3 | Center 4 | Center 5 | Center 6 | Center 7 | Center 8 |
|---|---|---|---|---|---|---|---|---|
| Total dose (GyRBE) | 36.25-40 | N/A | 36.25 | 36.25-40 | 35-38 | 35-40[1] | 38 | 40 |
| Fractions | 5 | 5 | 5 | 5 | 5 | 5 | 5 | 5 |
| Tx frequency | Every other day | Every other day | Every other day | Every other day | Every other day | Every other day | Every other day | Every other day |
| MRI for target contour | Yes | Yes | Yes | Sometimes | Yes | Yes | Yes | Yes |
| PTV for plan optimization | Yes | No | Yes | No | Yes | No | Yes | Yes |
| PTV for plan evaluation | Yes | No | Yes | No | Yes | No | Yes | Yes |
| Margins (CTV to PTV, mm) | 4-6 | N/A | 5 (except 3 in posterior) | N/A | 8 (L/R)[2]<br>3 (S/I)<br>3 (A/P) | N/A | 8 (L/R)[2]<br>3 (S/I)<br>3 (A/P) | 10 (L/R)[2]<br>5 (S/I)<br>5/4 (A/P) |
| Crop PTV from OAR | No | No | No | N/A | Yes[3] | N/A | No | Yes[3] |
| Beam Arrangements | 2 LATs & 2 ANT obliques[4] | 2 LATs | 2 LATs & 2 ANT obliques[4] | 2 LATs | 2 LATs | 2 LATs | 2 LATs | 2 LATs |

[1]Most typically used 40 GyRBE. [2]Range uncertainty is included in the margin and the structure is called PTV_eval. [3]Only when the OAR constraints cannot be met. [4]Beam weighs of lateral beams (70%) and anterior obliques beams (30%).

Table 3. Treatment plan optimization and evaluation for PBS.

| | Center 1 | Center 2 | Center 3 | Center 4 | Center 5 | Center 6 | Center 7 |
|---|---|---|---|---|---|---|---|
| Optimization Method | SFO/MFO | SFO | SFO/MFO | SFO | SFO | SFO | SFO |
| ROpt of target | Always | Always | Always | Always | Sometimes | Always | Never |
| ROpt of OAR | Yes | Yes | Yes | Yes | No | Sometimes[1] | No |
| Target for ROpt | CTV | CTV | CTV | CTV | CTV | CTV | N/A |
| ROpt setup margin (mm) | 3 | 5 (except 3 in posterior) | 5 (except 3 in posterior) | 5 | 3 | 3 | N/A |
| ROpt range uncertainty | 3.0% | 3.5% | 3.5% | 3.5% | 3.0% | 3.5% | N/A |
| REval used | Yes | Yes | Yes | Yes | Yes | Yes | No |
| Target for REval | CTV | CTV | CTV | CTV | CTV | CTV | N/A |
| REval setup margin | 3 mm | 5 mm | 5 mm | 5 mm | 3 mm | 3 mm | N/A |
| REval range uncertainty | 3.0% | 3.5% | 3.5% | 3.5% | 3.0% | 3.5% | N/A |
| REval scenarios | Worst case | Worst case | Worst case | N/A | Worst case | Worst case | N/A |
| REval target criteria | $D_{98\%}$>95% | $D_{98\%}$>95% | $V_{100\%}$> 95% | $D_{95\%}$>95% | $D_{98\%}$>99% | N/A | N/A |

*Abbreviations*: SFO for single field optimization, MFO for multi-fields optimization, ROpt for robust optimization, REval for robustness evaluation.

[1]Robust optimization is applied when the robust evaluation for the worst case does not meet the criteria for the target coverage.

Table 4. Dose Calculation, 2nd MU check and patient-specific QA.

| | Center 1 | Center 2 | Center 3 | Center 4 | Center 5 | Center 6 | Center 7 | Center 8 |
|---|---|---|---|---|---|---|---|---|
| Dose Calculation Algorithm | Analytical | Monte Carlo | Monte Carlo | Monte Carlo | Analytical | Analytical | Monte Carlo | Analytical |
| Statistics for Monte Carlo | N/A | 2% | 0.5% | 0.5% | N/A | N/A | 1% | N/A |
| Grid Size for analytical (mm) | 2 | N/A | N/A | N/A | 2.5 | 2.5 | 2.5 | 2 |
| 2nd MU check | Monte Carlo | No | Analytical | Analytical | Monte Carlo | Monte Carlo | No | No |
| PSQA Method | Logfile[1] | Measure | Measure | Measure | Measure | Measure | Measure | Measure |
| Measured Planes | 2 | 3 | 1 | 2 | 1 | 2 | 1 | 1 |
| Log file analysis | Yes | N/A | N/A | N/A | Yes | N/A | N/A | N/A |
| PSQA Criteria | 3%/3mm, Gamma ≥95% | 3%/3mm, Gamma ≥95% | 2%/2mm, Gamma ≥90% | 3%/3mm, Gamma ≥90% | 3%/2mm, Gamma ≥95% | 3%/3mm, Gamma ≥90% | 3%/2mm, Gamma ≥95% | Output difference ≤2% |

[1]Logfile QA was applied to all patients, and measurements were only applied to randomly selected patients.

Table 5. IGRT method for patient alignment and verification CT for dose evaluation.

| | Center 1 | Center 2 | Center 3 | Center 4 | Center 5 | Center 6 | Center 7 | Center 8 |
|---|---|---|---|---|---|---|---|---|
| Rescanning | No | No | No | No | No | No | No | No |
| Structure for IGRT | Fiducial contour | Fiducial contour | Fiducial contour | Fiducial contour | Fiducial contour | Isodose Line ROI | Fiducial contour | Fiducial contour |
| IGRT type | kV/kV | CBCT | CBCT | kV/kV and CBCT | kV/kV | kV/kV and CBCT | kV/kV | kV/kV |
| kV per field | Yes | No | No | No | Yes | Yes | Yes | Yes |
| Post Tx kV | No | No | No | No | No | No | Yes | No |
| IGRT tolerance | 3 mm | 1 mm | 3 mm | 2 mm | 3 mm | 2 mm | 1 mm | 1 mm |
| Verification CT | Sometimes | Yes | No | Sometimes | No | Sometimes | No | Yes |
| Evaluate Dose | Sometimes | No | Sometimes | Yes | No | Sometimes | Sometimes | Sometimes |

Supplementary Table 1. Dose constraints for the prostate SBRT treatment by protons.

| | Center 1 | Center 2 | Center 3 | Center 4 | Center 5 | Center 6 | Center 7 | Center 8 |
|---|---|---|---|---|---|---|---|---|
| CTV constraints | $D_{98\%}$>100% | N/A | $V_{100\%}$>98% | $D_{95\%}$>100% | $D_{100\%}$>99% | $D_{95\%}$>100% | N/A | N/A |
| PTV constraints | $V_{99\%}$>98% | N/A | $D_{95\%}$>95% | N/A | $D_{95\%}$=100% | N/A | $D_{95\%}$=100% | $D_{95\%}$>95% |
| Rectum dose constraints | $D_{0.03cc}$< 40Gy | $D_{0.03cc}$< 38Gy | $D_{0.03cc}$< 38Gy | $D_{0.03cc}$< 38Gy | $D_{0.03cc}$< 105% | $D_{1cc}$<36Gy $D_{max}$<41.2Gy | $D_{1cc}$<38.9Gy | $D_{0.03cc}$<40Gy |
| Bladder dose constraints | $D_{0.03cc}$< 40Gy | $D_{0.03cc}$< 38Gy | $D_{0.03cc}$< 38Gy | $D_{0.03cc}$< 38Gy | $D_{0.03cc}$< 105% | $D_{max}$<41.2Gy | $D_{8cc}$<39Gy | $D_{0.03cc}$<40Gy |
| Bowel dose constraints | $D_{0.03cc}$< 35Gy | N/A | $D_{0.03cc}$< 38Gy[1] | N/A | $D_{0.03cc}$< 55Gy EQD2 | $D_{max}$< 25Gy | $D_{1cc}$<28.8Gy | $D_{0.03cc}$<30Gy |
| Femoral heads dose constraints | $D_{1cc}$< 50% | N/A | $V_{15.6Gy}$< 10cc | N/A | $D_{1cc}$< 50% | $D_{max}$< 25Gy | $D_{1cc}$<23Gy | $D_{1cc}$< 50% |

[1]Small bowel dose constraints were reported except center 3 reported large bowel dose constraints.

Appendix 1 – Survey Questions

| Question No. | Question |
|---|---|
| 1 | Institution Name |
| 2 | Proton system vendor |
| 3 | Beam Delivery Mechanism |
| 4 | Treatment planning system |
| | **Patient selection** |
| 5a | Does your clinic offer (to eligible patients) proton SBRT (fx ≤ 5) for prostate radiotherapy? |
| 5b | If no, what's the biggest factor preventing your center from performing prostate proton SBRT? |
| 5c | If yes, have you completed the corresponding IROC proton prostate phantom credentialing? |
| 5d | If yes, up to what risk level do you offer prostate SBRT? |
| 6 | Why would you choose SBRT for primary prostate cancer? |
| 7 | Why would you choose SBRT for prostate metastatic treatment? |
| 8 | Do you perform SBRT reirradiation? |
| 9 | What is the typical treatment intent? |
| 10 | What kind of palliative cases are treated? |
| 11 | Are patients with metal hip implants eligible for prostate SBRT with proton therapy? |
| | **Simulation** |
| 12 | What immobilization devices are used? |
| 13a | Is an endorectal balloon used? |
| 13b | If yes, what it is filled with? |
| 14 | Is a hydrogel spacer used? |
| 15a | Are fiducial markers used in the prostate? |
| 15b | If yes, what material are the fiducials? |
| 16 | What CT scan slice thickness is typically used for prostate SBRT? |
| | **Treatment planning and QA** |
| 17 | Do you use MRI to delineate the prostate and seminal vesicles? |
| | When using SBRT, what is your prescription? |
| 18a | Target |
| 18b | Total Dose |
| 18c | Volume (%) |
| 18d | # Fractions |
| 19 | Is a PTV used for plan optimization? |
| 20 | Is a PTV used for plan evaluation? |
| | If a PTV is used for plan optimization/evaluation, what CTV-PTV expansions are used? Please comment |
| 21a | Left/Right |
| 21b | Superior/Inferior |
| 21c | Anterior |
| 21d | When they overlap, do you crop the PTV to protect OARs (bladder, rectum)? |
| 22 | What is the frequency of treatment? |
| 23 | What is the typical beam arrangement for planning? |
| | What are the dosimetry parameters for the following targets? |
| 24a | PTV |
| 24b | CTV |
| 24c | GTV |
| 25a | Do you use analytical algorithm or MC for planning dose calculation ? |
| 25b | If you use analytical TPS, what grid size is used for dose calculation? |
| 25c | If you use MC TPS, what is the dose statistics error? |
| 26 | If you have pencil beam scanning, do you use single-field or multi-field optimization? |
| 27 | Do you use robust optimization for the prostate? |
| 28 | Do you use robust optimization on OARs? |
| 29 | On which targets is robust optimization utilized? |
| 30 | What robust optimization **setup** uncertainty is used? |
| 31 | What robust optimization **range** uncertainty is used? |
| 32 | Do you use robust evaluation for plan evaluation? |
| 33 | On which targets is robust evaluation performed? |
| 34 | What robust **evaluation** setup uncertainty is used? |
| 35 | What robust **evaluation** range uncertainty is used? |
| 36 | What is robust criterion for target and OARs? |
| 37a | What is the dose volume criterion for target evaluation for robustness analysis? |
| 37b | How many scenarios have to meet the above criteria? |
| 38 | Which target is used for coverage evaluation for the nominal plan? |
| 39 | What is the allowable dose (D0.03 cc) to the **rectum** for 5 fx SBRT? (if applicable) |
| 40 | What is the allowable dose (D0.03 cc) to the **bladder** for 5 fx SBRT? (if applicable) |
| 41 | What is the allowable dose (D0.03 cc) to the **small bowel** for 5 fx SBRT? (if applicable) |
| 42 | What is the allowable dose (D1cc[%]) to the **femoral heads** for 5 fx SBRT? (if applicable) |
| 43a | Do you perform 2nd dose calculation check? |
| 43b | If yes, is the 2nd check algorithm analytical or MC? |
| 44a | What kind of patient QA is performed? |
| 44b | For patient-specific QA, how many depths are measured for a typical prostate SBRT field? |
| 44c | If you use log file analysis, do you perform this QA prior to patient treatment, or use the 1st fx for QA? |
| 45 | What is the patient QA analysis criteria? |
| | |
| | **Motion Management and IGRT** |
| 46a | Is rescanning used? |
| 46b | If rescanning is used, is it in-layer or volumetric? |
| 47a | Are structures used for IGRT alignment? |
| 47b | If yes, what types of structures? |
| 48 | What types of IGRT are used? |
| 49 | Are kV images taken for each field? |
| 50a | Are post treatment images acquired? |
| 50b | If so, what imaging technique is used? |
| 52 | What is the tolerance for fiducial markers/contours/bony anatomy? |
| 53 | Do you do an evaluation or QA CT prior to the 1st treatment? |
| 54 | Is any dose recalculation performed during the course of treatment? |

# Appendix 2 – Follow-up mini survey questions

General question to all centers that answered "no SBRT for prostate in proton therapy":
1. Do you perform prostate SBRT with X-rays? If you do, what specific concerns do you have about not using protons for prostate SBRT?
2. If you perform standard or hypofractionated proton prostate treatment, is the lower reimbursement for proton prostate SBRT compared to fractionated proton prostate therapy a factor in your decision not to use protons for prostate SBRT?

Specific to the 4 centers who answered "no clinic need":
3.(A) You answered that the lack of clinic need is the reason for not using protons for prostate SBRT. If you perform standard fractionated or moderated hypofractionated proton treatment, do you believe there is an additional advantage above standard/moderate hypofractionation to deliver proton SBRT to reduce patient burden and improve resource utilization?

Specific to the 6 centers who answered "hardware limitation":
3.(B) You answered that the lack of volumetric imaging in the proton facility is the reason for not using protons for prostate SBRT. What is the reason you do not prefer to deliver proton SBRT for prostate with kV radiographs and fiducials?

Specific to the 7 centers who answered "lack of clinical evidence" and concern about clinic outcomes:
3.(C) You answered that the perceived lack of clinical evidence is the reason for not using protons for prostate SBRT. Would you consider starting to deliver proton SBRT as recent phase III studies for proton- and x-ray-based SBRT have been or will soon be published?